\documentclass[a4paper,aps,twocolumn,prl,
showpacs,amsmath,amssymb,superscriptaddress]{revtex4-1}

\usepackage{graphicx}
\usepackage{dcolumn}
\usepackage{bm}
\usepackage{longtable}
\usepackage{color}

\makeatletter
\newcommand\figcaption{\def\@captype{figure}\caption}
\newcommand\tabcaption{\def\@captype{table}\caption}

\makeatother

\begin{document}


\title{Direct $k$-space mapping of the electronic structure in an oxide-oxide interface}

\author{G. Berner}
\author{M. Sing}
\affiliation{Physikalisches Institut and R\"ontgen Center for Complex Materials
Systems (RCCM), Universit\"at W\"urzburg, Am
Hubland, D-97074 W\"urzburg, Germany}
\author{H. Fujiwara}
\affiliation{Division of Materials Physics, Graduate School of Engineering Science, Osaka University, Osaka 560-8531, Japan}
\author{A. Yasui}
\author{Y. Saitoh}
\affiliation{Condensed Matter Science Division, Japan Atomic Energy Agency, SPring-8, Hyogo 679-5148, Japan}
\author{A. Yamasaki}
\author{Y. Nishitani}
\affiliation{Faculty of Science and Engineering, Konan University, Kobe 658-8501, Japan}
\author{A. Sekiyama}
\affiliation{Division of Materials Physics, Graduate School of Engineering
Science, Osaka University, Osaka 560-8531, Japan}
\author{N. Pavlenko}
\affiliation{Center for Electronic Correlations and Magnetism, Experimental Physics VI, Universit\"at Augsburg, D-86135 Augsburg, Germany}
\affiliation{Center for Electronic Correlations and Magnetism, Theoretical Physics III, Universit\"at Augsburg, D-86135 Augsburg, Germany}
\affiliation{Max Planck Institute for Solid State Research,
Heisenbergstra{\ss}e 1, D-70569 Stuttgart, Germany}
\author{T. Kopp}
\affiliation{Center for Electronic Correlations and Magnetism, Experimental Physics VI, Universit\"at Augsburg, D-86135 Augsburg, Germany}
\author{C. Richter}
\affiliation{Center for Electronic Correlations and Magnetism, Experimental Physics VI, Universit\"at Augsburg, D-86135 Augsburg, Germany}
\affiliation{Max Planck Institute for Solid State Research, Heisenbergstra{\ss}e
1, D-70569 Stuttgart, Germany}
\author{J. Mannhart}
\affiliation{Max Planck Institute for Solid State Research, Heisenbergstra{\ss}e
1, D-70569 Stuttgart, Germany}
\author{S. Suga}
\affiliation{Institute of Scientific \& Industrial Research, Osaka University, Ibaraki, Osaka 567-0047, Japan}
\author{R. Claessen}
\affiliation{Physikalisches Institut, Universit\"at W\"urzburg, Am Hubland,
D-97074 W\"urzburg, Germany}

\date{\today}

\begin{abstract}
The interface between LaAlO$_3$ and SrTiO$_3$ hosts a two-dimensional electron
system of {\it itinerant} carriers, although both oxides are band insulators.
Interface ferromagnetism coexisting with superconductivity has been found and
attributed to {\it local} moments. Experimentally, it has been established that
Ti~$3d$ electrons are confined to the interface. Using soft x-ray angle-resolved
resonant photoelectron spectroscopy we have directly mapped the interface states
in $k$-space. Our data demonstrate a charge dichotomy. A mobile fraction
contributes to Fermi surface sheets, whereas a localized portion at higher
binding energies is tentatively attributed to electrons trapped by O-vacancies in
the SrTiO$_3$. While photovoltage effects in the polar LaAlO$_3$ layers cannot
be excluded, the apparent absence of {\it surface-related} Fermi surface sheets
could
also be fully reconciled in a recently proposed electronic reconstruction
picture where the built-in potential in the LaAlO$_3$ is compensated by
{\it surface} O-vacancies serving also as charge reservoir.
\end{abstract}

\maketitle

Breaking the translational or inversion symmetry at surfaces and interfaces may
lead to a rearrangement of charge, spin, orbital, and lattice degrees of
freedom. The consequences are particularly interesting in the case of oxides,
where already a slight shift in the balance of the respective interactions
can stabilize one out of several competing orders or even create novel phases. The case
in hand is the formation of a high-mobility two-dimensional electron
system (2DES) from Ti~$3d$ states at the interface of LaAlO$_3$/SrTiO$_3$
(LAO/STO) heterostructures \cite{Ohtomo04, Thiel06, Nakagawa06, Reyren07,
Basletic08, Sing09, Salluzzo09, Drera11, Koitzsch11} which undergoes a
transition into a two-dimensional superconducting state below
0.2\,K \cite{Reyren07}. However, depending on growth conditions LAO/STO has also
been found to display pronounced magnetotransport effects indicating the
existence of local moments \cite{Brinkman07}. More recently, even the
simultaneous presence of ferromagnetism and superconductivity has been reported,
possibly due to phase separation within the interface \cite{Li11, Bert11,
Kalisky12, Pavlenko12}.

The physical origin of the 2DES formation is still debated. The observation that
both interface conductivity as well as ferromagnetism only appear for a critical
LAO thickness of 4 unit cells (uc) and beyond has been related to electronic
reconstruction \cite{Thiel06, Kalisky12}. In this scenario electrons are
transferred from the surface to the interface in order to minimize the
electrostatic energy resulting from the polar discontinuity between LAO and
STO \cite{Nakagawa06}. Alternative explanations involve doping by oxygen
vacancies \cite{Kalabukhov07,Siemons07} and/or cation intermixing
\cite{Willmott07} but so far
have failed to account for the critical thickness. What hampers a better
understanding is the lack of microscopic information, in particular on
the electronic properties of the interface and the adjacent oxide layers.

Photoelectron spectroscopy is unique in that it can directly probe the
single-particle excitations of the valence electrons, which determine the
low-energy properties of a solid. Regarding the Ti~$3d$ interface electrons in
LAO/STO, however, this is hindered for conventional photon energies in the range
of 20--100\,eV by the insufficient probing depth while in the hard x-ray regime
the photoionization cross-sections are too low. Only recently, it has been shown
that the Ti~$3d$ states can be detected by exploiting the resonance
enhancement at the Ti~$L$ edge, i.e. by utilizing soft x-rays \cite{Drera11,
Koitzsch11}. Here we use soft x-ray resonant photoelectron spectroscopy
(SX-ResPES) to probe the occupied part of the electronic states and in
particular its angle-resolved (AR) mode to record Fermi surface (FS) and band
maps. Note that the resulting electronic structure may differ from that probed
in transport measurements due the additional presence of x-ray induced
photocarriers \cite{footnote1,Sing09}.

The LAO/STO heterostructure with a 4\,uc thick LAO overlayer was grown at the
University of Augsburg by pulsed laser deposition on a TiO$_2$-terminated STO
substrate. The film thickness was monitored by reflection high-energy electron
diffraction (RHEED). During growth the oxygen pressure amounted to $1 \times
10^{-4}$\,mbar while the substrate was held at 780\,$^\circ$C. Subsequently, the
sample was cooled down to room temperature in 0.4\,bar of oxygen. Prior to
the measurements, the sample surface was cleaned by ozone and gentle {\it
in situ} heating at 180$^{\circ}$C \cite{SupplMat3}.
Note that for our samples it has been shown previously that the interface
becomes conducting only if a critical thickness of 4\,uc is reached
\cite{Thiel06} and that the highly mobile electrons \cite{Schneider06} are
confined within a few unit cells on the STO side of the interface where they
occupy Ti~$3d$ states \cite{Sing09, Berner10, Zhou11}.

The experiments were performed at the soft X-ray beamline BL23SU
of SPring-8 using a photoemission spectrometer equipped with a Gammadata-Scienta
SES-2002 electron analyzer \cite{Saitoh12} and the fully circularly polarized
light from a helical undulator. The energy resolution was set to 110\,meV for
the angle-integrated and 180\,meV for the angle-resolved experiments while the
sample temperature was 20\,K for all measurements. The angular resolution of
the SX-ARPES measurements was 0.2$^{\circ}$ along and 0.5$^{\circ}$
perpendicular to the analyzer slit. The Fermi surface map was generated by
integrating energy distribution curves at each $k$-point over an interval of
0.3\,eV centered around the Fermi energy. All spectra were
corrected for the contribution from second order light \cite{SupplMat2}. The
position of the Fermi level was determined from an {\it in situ} evaporated gold
film \cite{SupplMat1}.

The density functional calculations have been performed using the generalized
gradient approximation (GGA+$U$) in the Perdew-Burke-Ernzerhof pseudopotential
implementation \cite{Perdew96} in the QUANTUM ESPRESSO (QE)
package \cite{Giannozzi09}, with the local Coulomb repulsion $U$ between Ti~$3d$
electrons being 2\,eV \cite{SupplMat4}.

\begin{figure}
\includegraphics[width = 0.48\textwidth]{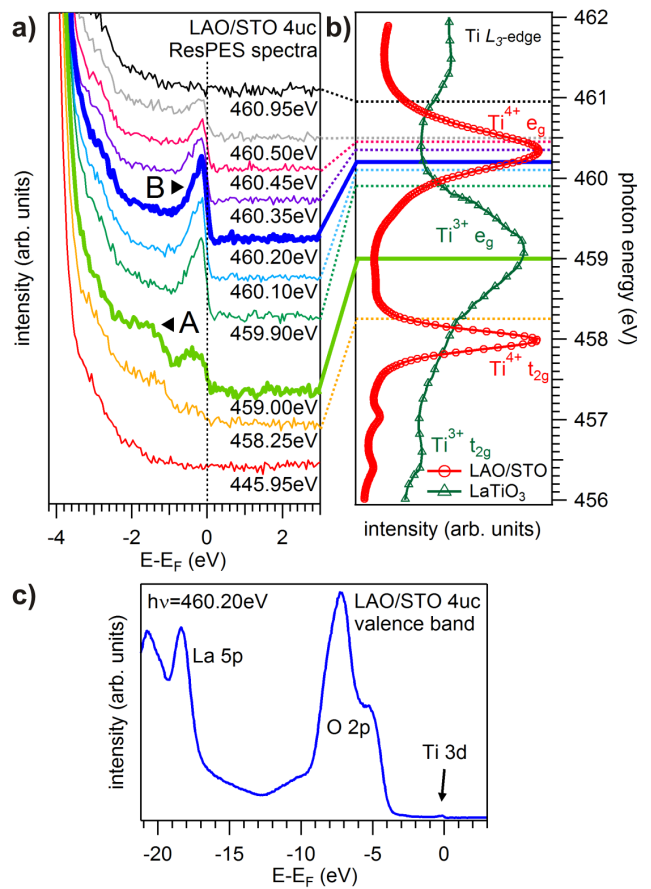}
\caption{\label{Fig:ResPes} (Color online) (a) Angle-integrated
resonant photoemission spectra upon
tuning the photon energy through the Ti~$L$ absorption edge. (b) Ti~$L$
absorption spectra, recorded in total electron yield mode, of LAO/STO (red) and
LaTiO$_3$
(green, taken from Ref.~\cite{Salluzzo09}). The photon energies used in the
measurements displayed in (a) are indicated by dashed and solid lines. (c)
On-resonance photoemission spectrum, displaying the complete valence band.}
\end{figure}

\begin{figure}
\includegraphics[width = 0.48\textwidth]{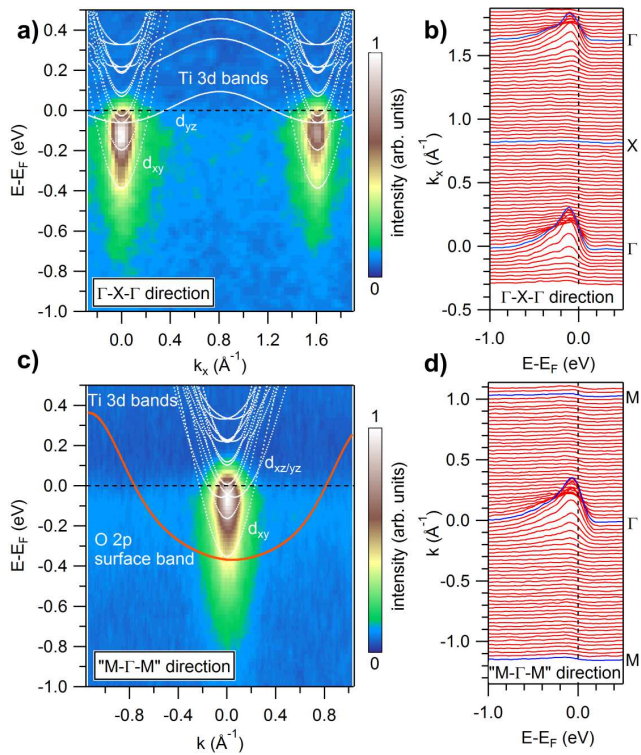}
\caption{\label{Fig:Arpes} (Color online) (a) Band map
along the $\Gamma$-X-$\Gamma$ line
of the BZ. (b) Same data as in (a) depicted as energy distribution curves. (c)
Band map along a cut close to the M-$\Gamma$-M line of the BZ (hence denoted by
"M-$\Gamma$-M"). (d) Same data as in (c), depicted as energy distribution
curves. All data were taken at $h\nu = 460.20$\,eV. White and red lines
represent theoretical band dispersions (for details see text).}
\end{figure}

\begin{figure*}
\includegraphics[width = 0.9\textwidth]{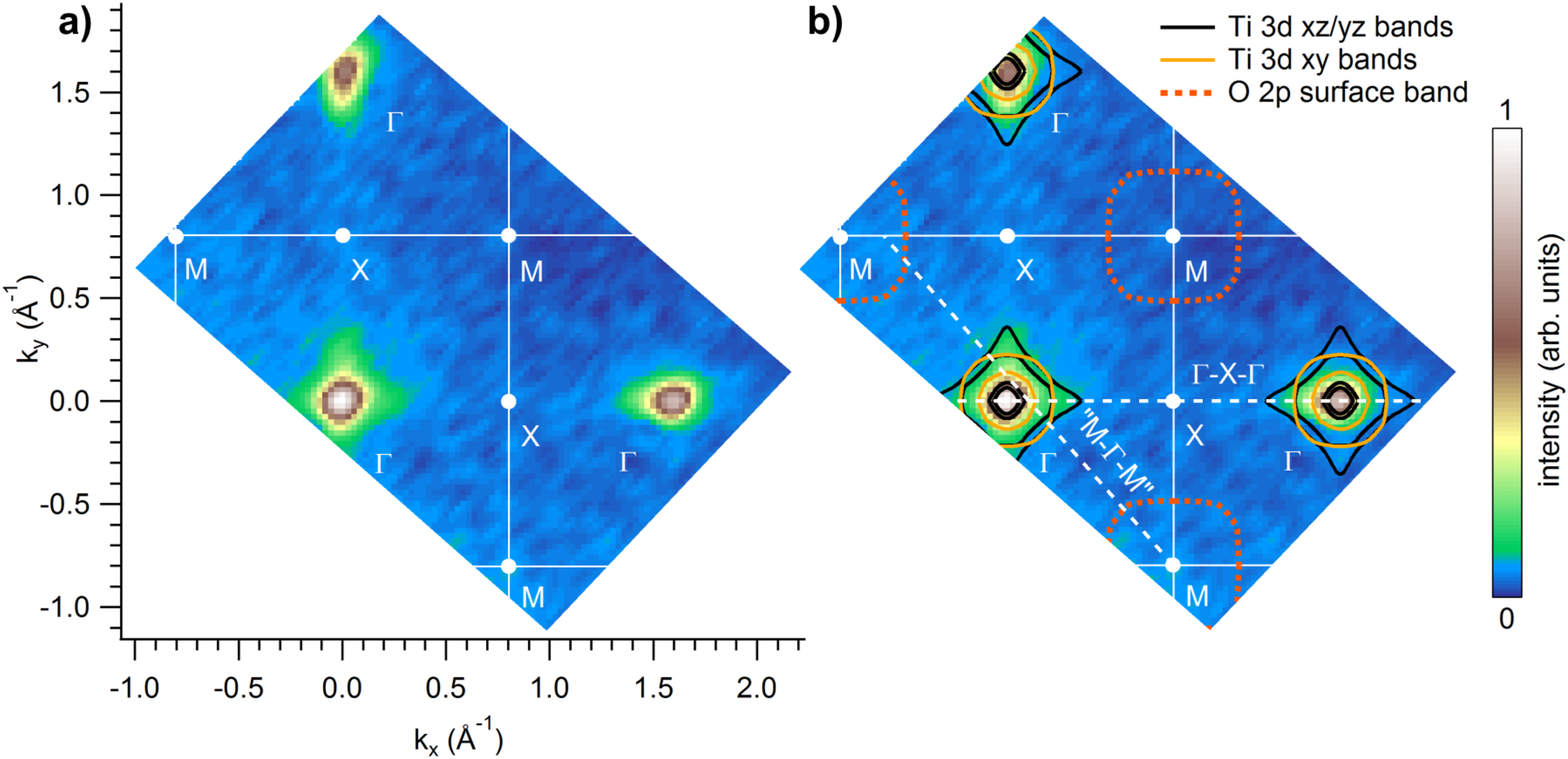}
\caption{\label{Fig:FSmap} (Color online) (a) FS map recorded at $h\nu =
460.20$\,eV. (b) Same map as (b) but with the cuts in the BZ corresponding to
the data in Fig.~\ref{Fig:Arpes}(a)-(d) indicated and the FS sheets from DFT
calculations overlaid.}
\end{figure*}

In Fig.~\ref{Fig:ResPes}(a) we show angle-integrated ResPES spectra near the
chemical potential for a sample with a conducting interface (4uc LAO) upon
tuning the photon energy through the Ti~$L$ absorption edge. Exciting at a
specific absorption edge makes ResPES element-specific. Two processes with
the same final state quantum mechanically interfere, namely the
direct photoemission and the (coherent and element-specific) Auger emission of
the outer valence electrons (involving the direct recombination of a valence
electron with the core hole). Thus, any enhancement of spectral weight upon
tuning the energy through an absorption edge corresponds to the valence
electrons of the respective atomic species, in our case Ti~$3d$ states (for
details on ResPES see Supplemental Material, Section I \cite{SupplMat1}).

The complete valence band spectrum, measured on
resonance ($h\nu$=460.20\,eV), is depicted in Fig.~\ref{Fig:ResPes}(c). The
off-resonance spectrum ($h\nu$=445.95\,eV) shows no spectral weight at the
chemical potential. Moving through the resonance two structures appear ---
indicating that both are of Ti~$3d$ character ---, a broad structure at a
binding energy of $\approx 1.3$\,eV ($A$) and a structure which is cut off by
the Fermi-Dirac distribution function [marked by $B$ in Fig.~\ref{Fig:ResPes}
(a)] and hence is tentatively ascribed to metallic
states. Interestingly, the two features $A$ and $B$ resonate at different
photon energies which already signals that they originate from different types
of electronic states.

To get further insight, one has to relate the ResPES excitation energies to
their positions on the resonance curve, i.e. the x-ray absorption spectrum
[Fig.~\ref{Fig:ResPes}(b)]. The spectrum to compare with is that of a reference
sample representative for Ti in a 3+ oxidation state, here LaTiO$_3$ (green),
while for LAO/STO (red) the absorption is dominated by the Ti$^{4+}$ ions of the
substrate. Peak $A$ resonates exactly on the absorption maximum (associated with
the so-called e$_g$ levels) of LaTiO$_3$ (Ti$^{3+}$) whereas the maximum enhancement of
$B$ is delayed by $\approx 1$\,eV to almost the following absorption minimum. In addition,
feature $B$ resonates over a wider energy range than feature $A$. This
phenomenology is known from ResPES on transition metals \cite{Kaurila97} and
indicates that features $A$ and $B$ should originate from localized and
delocalized states, respectively \cite{SupplMat1}.
While the delocalized states ($B$) are readily identified as Ti~$3d$ band
states, forming the 2DES, we assign the localized states ($A$) to Ti~$3d$ impurity states. Such
trapped states are likely to be induced by O vacancies adjacent to Ti~ions. Our
recent band structure calculations based
on density-functional theory (DFT) reveal similar peaks in the density of states
between $-2$\,eV and $-1$\,eV, depending on O vacancy configuration and
concentration \cite{Pavlenko12}. Similar observations have been reported for
scanning tunneling spectroscopy measurements on LAO/STO \cite{Ristic12} and PES
experiments on bare STO surfaces \cite{Aiura02, Ishida08}, where the spectral
weight at the chemical potential and the in-gap states around $-1.3$\,eV were
discussed in terms of coherently screened and poorly screened
excitations \cite{Ishida08}, respectively.

Since momentum information is still preserved in valence band photoemission
using soft x-rays and due to the resonance enhancement at the Ti~$L$ edge we
have been able to perform k-space mapping of the 2DES interface states
[Figs.~\ref{Fig:Arpes}(a) and (c)]. In Figs.~\ref{Fig:Arpes}(b) and (d) we
display the corresponding $k$-resolved energy dispersion curves along the
$\Gamma$-X-$\Gamma$ direction and a cut close to the M-$\Gamma$-M line [see
dashed lines in Fig.~\ref{Fig:FSmap}(b)]. One clearly observes states
dispersing with $k$ around $\Gamma$ and an occupied band width of $\approx
0.4$\,eV, conclusively confirming our tentative assignment from above to the
metallic interface band states of the 2DES. In Figs.~\ref{Fig:Arpes}(a) and (c),
the electronic dispersions of DFT calculations are overlaid. The calculations
fairly reproduce the experimental band width, while the experimental broadening
does not allow us to resolve all the individual quantum well states owing to the
confinement of the 2DES \cite{Santander-Syro11}, although the energy
distribution curves in Figs.~\ref{Fig:Arpes}(b) and (d) indicate the existence
of at least two bands.

\begin{figure}
\includegraphics[width = 0.48\textwidth]{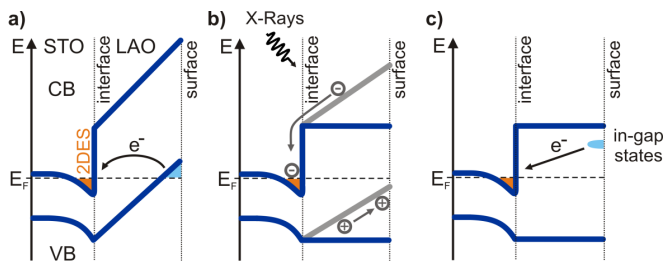}
\caption{\label{Fig:BandDiagrams} (Color online) (a) Schematic band diagram for
the standard electronic reconstruction scenario in LAO/STO. (b) Possible
situation where the polarization field in the LAO is screened by the separation
of electron-hole pairs created upon illumination with x-rays. (c) Tentative band
situation with the inclusion of O vacancies at the LAO surface as charge
reservoir for electronic reconstruction.}
\end{figure}

Additional information can be extracted from the FS map in
Fig.~\ref{Fig:FSmap}. An essentially isotropic distribution of high
intensity
centered at the $\Gamma$ points of the BZ of STO is observed. Superimposed one
finds, with much lower intensity, flower-shaped spectral weight with the lobes
directed towards the X points of the BZ and stretching further out than the
isotropic intensity distribution as is most clearly seen around the lower left
$\Gamma$ point in Fig.~\ref{Fig:FSmap}(a). These observations are in line with
the FS sheets from the DFT calculations which are overlaid on top of the PES
data in Fig.~\ref{Fig:FSmap}(b). Note, however, that the experimental Fermi
surface volume might be slightly enhanced due to photo-generated charge
carriers \cite{Sing09}. The calculations further reveal that the isotropic
intensity distribution originates mainly from light Ti~$3d_{xy}$ bands while the
flower-shaped intensity distribution is due to heavy $d_{xz}$/$d_{yz}$ [Fig.
\ref{Fig:Arpes}] bands, similar to what has recently been reported for a 2DES at
the surface of bare STO \cite{Santander-Syro11}.

However, there is also a striking discrepancy between experiment and theory. In
the calculations, which are performed for stoichiometric samples, i.e., in the
absence of oxygen vacancies in the LAO, a hole-like FS is predicted around the M
points as indicated by the orange dashed lines in Fig.~\ref{Fig:FSmap}(b). It
is due to the Fermi level crossing of O~$2p$ derived states from the valence
band maximum of the topmost monolayer of the LAO film. Since the photoelectron
current emitted from the surface is hardly damped by scattering events and since
the band is almost completely filled, these hole pockets, if existing, should be
observable even without resonance enhancement. In the standard electronic
reconstruction scenario [Fig.~\ref{Fig:BandDiagrams}(a)] the polarity-induced
potential build-up across the overlayer gradually shifts the valence band of LAO
towards the Fermi level. At the critical thickness, the valence band maximum
crosses the Fermi level which gives rise to the hole pockets predicted by the
DFT calculations, while the released electrons populate the lowest lying Ti~$3d$
states at the STO side of the interface \cite{Pentcheva09}.

From the obvious absence of metallic surface bands we conclude that the
potential difference is (almost completely) screened out in the LAO film
\cite{footnote2}. A possible explanation
could be that under illumination electron-hole pairs are created which get
separated by the initial polarization field in the LAO
[Fig.~\ref{Fig:BandDiagrams}(b)]. Eventually, depending on the rates for
electron-hole pair creation and recombination a dynamic equilibrium will be
established where the opposing electric field due to the separated electron-hole
pairs just cancels the initial field, thus prohibiting the observation of
the surface related hole pocket. On the other hand, in another
polar/non-polar heterostructure (LaCrO$_3$/SrTiO$_3$) a built-in potential has
readily been identified by x-ray photoelectron spectroscopy (XPS) with a similar
potential drop per uc as predicted for LAO/STO in the standard electronic
reconstruction scenario \cite{Chambers11}.

Looking for alternative explanations, our SX-ARPES results could also be fully
reconciled within recent proposals, also based on DFT calculations, that oxygen
vacancies at the LAO surface can serve as charge reservoir for the
electronic reconstruction \cite{Zhong10, Bristowe11, Li11a, Pavlenko12a}. In
such a scenario the O vacancies induce unoccupied in-gap states, which due to
possible disorder of the vacancies in the real system might easily become
localized. The released two electrons per O vacancy are transferred to the
interface by the LAO polar field. This field is thereby efficiently reduced [see
Fig.~\ref{Fig:BandDiagrams}(c)], in line with XPS measurements which show no
core-level broadening or shift with film thickness \cite{Segal09}.

Viewing the LAO film as a parallel plate capacitor the energy gain per charge
transferred from the surface to the interface increases linearly with film
thickness. While the energy costs for the creation of oxygen vacancies can
depend also on thickness (and in addition on vertical position), the
calculations show that at a certain thickness it becomes on balance favorable to
create an oxygen vacancy at the surface and transfer the released electrons to
the interface \cite{Zhong10, Bristowe11, Li11a, Pavlenko12a}. Note that O
vacancies in the STO can be excluded with fair certainty as source of electrons
for the 2DES. First, their density essentially does not depend on film thickness
but on the growth conditions and so the critical thickness behavior as
observed in our samples could hardly be explained \cite{Thiel06}. Second, in
our data the electrons associated with O vacancies in the STO (feature $A$) are
well separated and far below the 2DES related states at the chemical potential
(feature $B$) and thus cannot populate the 2DES related states.

Summarizing our results we arrive at the picture that at the conducting
interface besides heavy Ti~$3d_{xz/yz}$ and light Ti~$3d_{xy}$ bands there also
exist \textit{localized} charge carriers of Ti~$3d$ character. These are
probably trapped by adjacent O vacancies, i.e. O vacancies in the STO at the
interface. It is hence tempting to associate the trapped and mobile interface
charge with ferromagnetism and superconductivity, respectively
\cite{Pavlenko12}, which both have recently been reported to
coexist at the interface. As a matter of fact, the hallmarks of the microscopic
view of the standard electronic reconstruction scenario --- a metallic surface
and a potential drop across the LAO overlayer of the order of the STO band gap
--- experimentally remained elusive so far. Our data
clearly indicate that they are absent in our spectroscopic experiments. Modified
electronic reconstruction scenarios involving surface O vacancies as charge
reservoir could explain these findings. Since we cannot rule out the possibility
that the LAO polar field is screened out by a photoinduced reverse voltage our
results call for both a systematic investigation of scenarios taking O vacancies
into account and combined \textit{in situ} transport and spectroscopy
experiments. In any case, these results demonstrate for the first time that
SX-ARPES can provide valuable $k$-space information on the electronic structure
of \textit{buried} interfaces.

\begin{acknowledgments}
We are grateful to T. Kiss for setting-up and optimizing the ARPES spectrometer
and C. Hughes for help with the sample preparation. This work was supported by
the Deutsche Forschungsgemeinschaft (FOR 1162 and TRR 80), the German Federal
Ministry for Education and Research (05 K10WW1) and by the Grant-in-Aid for
Innovative Areas (20102003) "Heavy Electrons" from MEXT, Japan. The measurements
were performed under the Shared Use Program of JAEA Facilities (Proposal No.
2011B-E31) and the approval of BL23SU at SPring-8 (Proposal No. 2011B3820). HF
was supported by MEXT/JSPS KAKENHI Grant Number 23740240.
\end{acknowledgments}

\clearpage

\begin{center}
{\Large \textsc{Supplemental Information}}
\end{center}

\section{Resonant photoelectron spectroscopy (ResPES) $ - $ technique and data normalization}

\begin{figure*}[h]
\includegraphics[width = 0.8\textwidth]{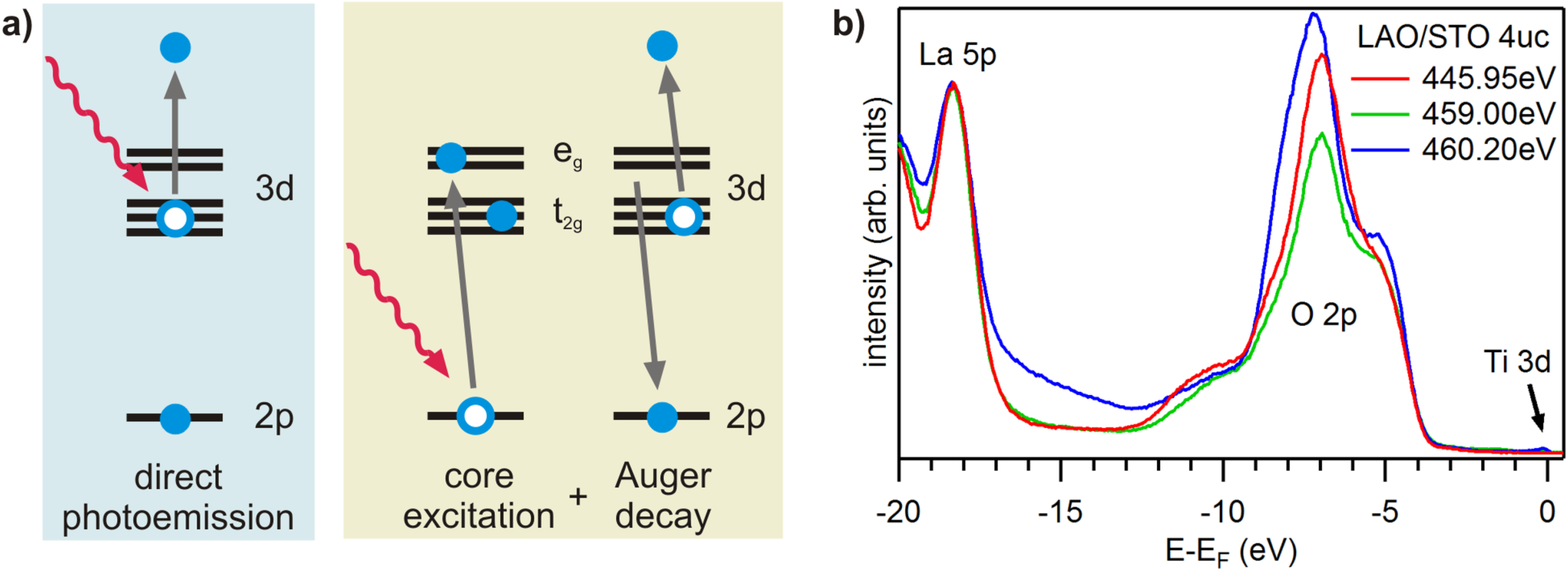}
\caption{\label{Fig:ResPesTech} (Color online) Resonant enhancement in
photoemission and data normalization. (a) Sketch of the two excitation channels
which quantum mechanically interfere in resonant photoemission. (b) The spectra
have been normalized to the non-resonating La~$5p$ core-level as is shown here
for spectra measured at $h\nu=445.95$\,eV (off-resonance), $h\nu=459.00$\,eV
(Ti$^{3+}$ e$_g$ resonance), and $h\nu=460.20$\,eV (on-resonance for Fermi
surface mapping). The resonance enhancement in the O~$2p$ region is due to the O~$2p$ - Ti~$3d$ hybridization.}
\end{figure*}

Photoemission (PES) from the Ti $3d$ interface states can only be accomplished
with reasonable intensity, if one takes advantage of the resonance enhancement
at the Ti $L$ absorption edge. With the photon energy tuned to the absorption
threshold, a second path opens how to arrive at the same final state as in the
direct photoemission process [Fig.~\ref{Fig:ResPesTech}(a)]:

\begin{gather}
2p^63d^n  \rightarrow 2p^63d^{n-1} + \epsilon \text{\qquad (direct PES),}\\
2p^63d^n \rightarrow  2p^53d^{n+1}  \rightarrow  2p^63d^{n-1} + \epsilon
\text{\qquad (Auger decay),}
\end{gather}
where $\epsilon$ denotes the ejected photoelectron. This second path involves
the dipole excitation of an electron from the Ti $2p$ into the Ti~$3d$ shell and
a subsequent Auger-like decay (often called direct recombination), resulting in
the ejection of a $3d$ electron and a filled $2p$ shell. The probability
amplitudes of both channels interfere quantum mechanically and give rise to a
resonance enhancement of only the $3d$ spectral weight \cite{Huefner03}. This picture is based on an atomic description of the 3d electrons. Note
that in the case of metallic (Bloch-like) electrons in a solid the simple
situation as sketched in
Fig.~\ref{Fig:ResPesTech}(a) might be refined. By hybridization, neighboring
atoms interact and the discrete atomic levels broaden into wide bands. In
ResPES, part of the energy of the initial photon now can be stored in a broad
continuum of excitations between occupied and unoccupied band states, leading to
a delayed resonance maximum and a stretched resonance range. Such a behavior is
known from transition metals \cite{Kaurila97S} and also observed here.

All spectra were normalized in intensity to the non-resonating La $5p$
core-level intensity at $-18.4$\,eV [see Fig.~\ref{Fig:ResPesTech}(b)]. The
energy scale was calibrated at the Fermi level as determined from a gold film
which has been freshly evaporated on the metallic sample carrier beside the sample after the measurements. By shorting the two-dimensional electron system
(2DES) and the grounded sample carrier with silver paint applied to the
side faces of the sample, we ensured that also the 2DES is in galvanic contact
with the spectrometer and hence 2DES and Au film share the same chemical
potential.

\section{Correction for 2$^{\text{nd}}$ order contributions}
Significant intensity from the Ti $2p_{3/2}$ core level excited by
2$^{\text{nd}}$ order light appeared in all spectra in the region around the
chemical potential [see Fig.~\ref{Fig:SecOrder}(a)] although the
2$^{\text{nd}}$ order light is much weaker than the intensity of the
1$^{\text{st}}$ harmonic due to the use of the circularly polarized light from a helical undulator.
For spectra measured at photon energies between
459.90\,eV and 460.95\,eV the 2$^{\text{nd}}$ order light induced peak was
fitted
by a Gaussian line and subtracted from both the angle-integrated and
angle-resolved spectra.

For the spectra measured at lower photon energies (445.95\,eV, 458.25\,eV and
459.00\,eV) the 2$^{\text{nd}}$ order peak overlaps with 1$^{\text{st}}$ order
spectral weight below the chemical potential. It was subtracted by using the
Gaussian lineshape observed at higher photon energies and accounting for the
$2\Delta h \nu$ shift of second order light, if the fundamental photon energy
changes by $\Delta h \nu$. The second order excitation of the Ti $2p$ electrons
also allowed for an absolute calibration of the photon energies used in
photoemission and x-ray absorption.

\section{Sample surface preparation}

\begin{figure*}[h]
\includegraphics[width = 0.8\textwidth]{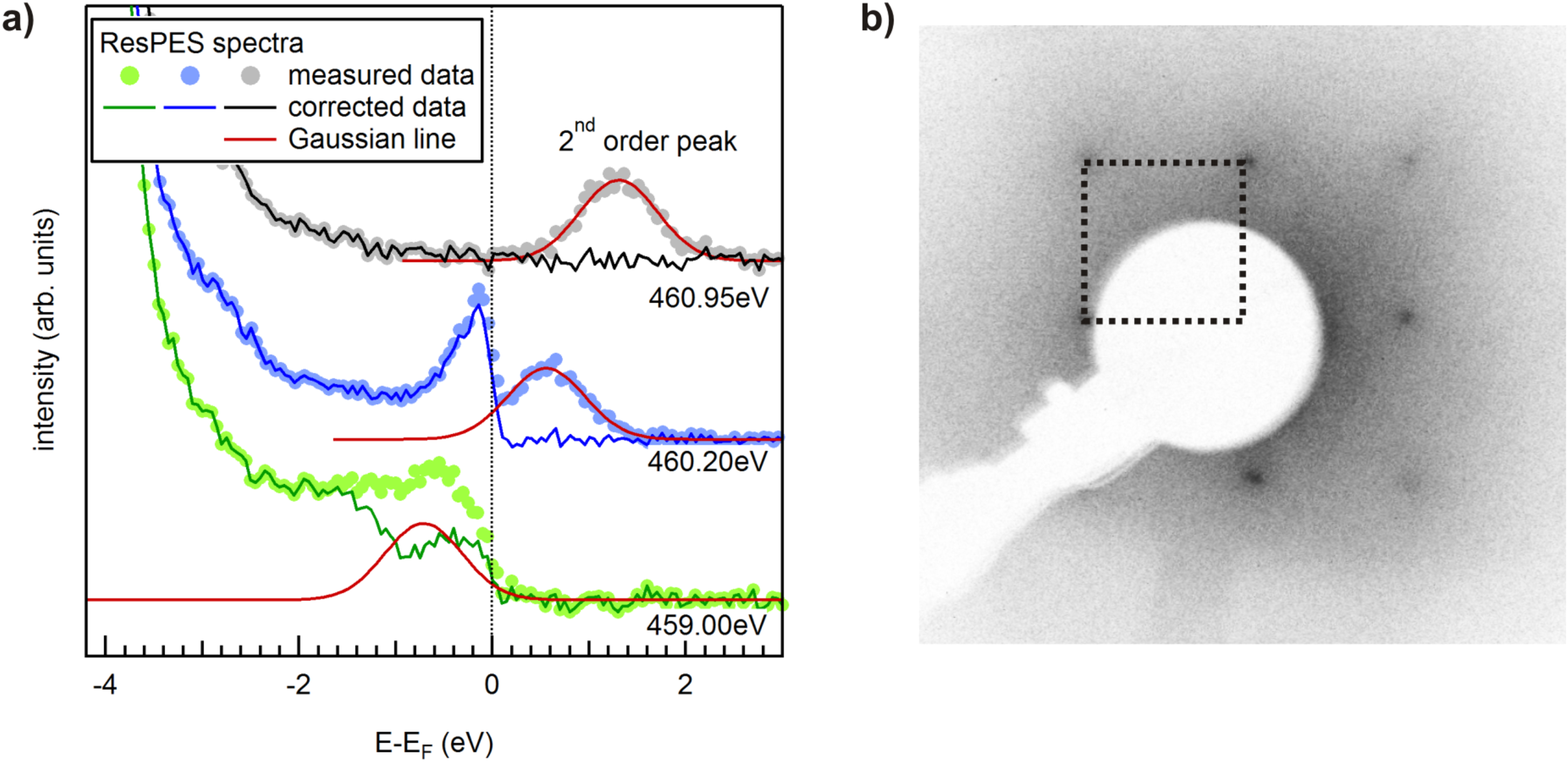}
\caption{\label{Fig:SecOrder} (Color online) (a) Correction of the 2$^\text{nd}$
order light induced peak by subtraction of a Gaussian, exemplarily shown for
spectra measured at 459.00\,eV, 460.20\,eV and 460.95\,eV. (b) LEED pattern of a
surface-cleaned LAO/STO sample at $E=128.1$\,eV showing a $1 \times 1$ surface
structure. The lattice constant is 3.97\,{\AA}.}
\end{figure*}

For the measurements a clean sample surface is essential to minimize scattering
of the photoelectrons and optimize the signal-to-noise ratio of the spectra. The
sample was transported under air to the synchrotron radiation laboratory. Prior to the measurements
the sample was
kept under ozone flow for 45\,min, followed by an {\it in situ} annealing at
180$^\circ$C under $1 \times
10^{-5}$\,mbar of oxygen for 45\,min.
After this procedure a high quality LEED pattern with a $1 \times 1$
surface is observed [see Fig.~\ref{Fig:SecOrder}(b)], signalling a
clean and long-range ordered LAO surface. The lattice constant inferred from
this pattern is $(3.97 \pm 0.1)$\,{\AA}, which is within the error bars in good
agreement with the lattice constant of STO (3.905\,{\AA}). No surface
reconstruction was found.

\section{Density functional calculations}

Density functional calculations have been performed using the generalized
gradient approximation (GGA+$U$) in the Perdew-Burke-Ernzerhof pseudopotential
implementation \cite{Perdew96S} in the QUANTUM ESPRESSO (QE)
package \cite{Giannozzi09S}. The local Coulomb repulsion $U$ between
Ti~$3d$ electrons was chosen to be 2\,eV.

A number of supercells consisting of two symmetric LAO/STO
parts were generated, where each part contains a stack of 4 uc thick LAO layers
on a 3.5 uc thick STO slab. The interfacial configuration is considered as
TiO$_2$/LaO. The LAO-STO-LAO parts are separated by a 13\,{\AA}-thick vacuum
sheet. A kinetic energy cutoff of 640\,eV and the Brillouin zone (BZ) of the
106-to 166-atom supercells sampled with $5 \times 5 \times 1$ to $9 \times 9
\times 1$ $k$-point grids are used. For the stoichiometric vacancy-free
structures, the electronic state can be characterized as metallic, with the
interface electron charge emerging due to the compensation of the polar field
across the LAO. The number of electrons per ($1 \times 1$) uc transferred to the
t$_{2g}$ states of the TiO$_2$-layers at the interface amounts to 0.34.

\begin{thebibliography}{30}
\bibitem{footnote2} The leading edge of the O~$2p$-derived states
(originating from both LAO and STO) in Fig.~\ref{Fig:ResPes}(b) indicates an
energy gap of about 3.2\,eV, corresponding within experimental accuracy to the
STO band gap. If there was a sizeable potential difference across the LAO
left the gap of the leading edge (then originating solely from LAO) would have
to be significantly reduced.
\end{thebibliography}
\end{document}